\documentclass[twocolumn,showpacs,preprintnumbers,amsmath,amssymb,nofootinbib,epsfig,bmamsfonts,yfonts]{revtex4}
\usepackage{graphicx}

\begin{document}

\title{The chiral magnetic wave in an expanding QCD fluid}


\author{Seyed Farid Taghavi$^{(1,2)*}$ and Urs Achim Wiedemann$^{(2)}$\footnote{E-mail: s.f.taghavi@ipm.ir, urs.wiedemann@cern.ch}}

\affiliation{$^{(1)}$School of Particles and Accelerators,
Institute for Research in Fundamental Sciences (IPM),
P.O.Box 19395-5531, Tehran, Iran\\
$^{(2)}$Physics Department, Theory Unit, CERN, CH-1211 Gen\`eve 23, Switzerland}

\begin{abstract}
As a consequence of the chiral anomaly, the hydrodynamics of hot QCD matter coupled to 
QED allows for a long-wavelength mode of chiral charge density, the chiral magnetic wave (CMW),
that provides for a mechanism of electric charge separation along the direction of an external
magnetic field. Here, we investigate the efficiency of this mechanism for values of the 
time-dependent magnetic field and of the energy density attained in the hot QCD
matter of ultra-relativistic heavy ion collisions. To this end, we derive 
the CMW equations of motion for expanding systems by treating the CMW
as a charge perturbation on top of an expanding Bjorken-type background field in the limit
$\mu/T \ll 1$. Both, approximate analytical and full numerical solutions to these equations
of motion indicate that for the lifetime and thermodynamic conditions of ultra-relativistic
heavy ion collisions, the efficiency of CMW-induced electric charge separation 
decreases with increasing center of mass energy and that the effect is numerically very
small. We note, however, that if sizable oriented asymmetries in the axial charge distribution 
(that are not induced by the CMW) are present in the early fluid dynamic evolution, then the 
mechanism of CMW-induced electric charge separation can be much more efficient. 
\end{abstract}


\maketitle

\section{Introduction.}
Our understanding of how the chiral anomaly 
affects the dynamics of relativistic plasmas has deepened 
considerably in recent years. First insights into macroscopic effects of
quantum anomalies go back to the early 1980s, when Vilenkin found
that rotating thermal matter emits equilibrium (anti)neutrino
currents that - as a consequence of their handedness - are oriented 
preferentially along (opposite to) the angular velocity 
vector~\cite{Vilenkin:1979ui,Vilenkin:1980zv} or an external 
magnetic field~\cite{Vilenkin:1980fu}. Related macroscopic 
parity-violating effects were discussed in the 1990s for systems
as different as superfluid  $^3$He-A~\cite{Volovik:1999wx}, and
the electro-weak plasma in the Early 
Universe~\cite{Joyce:1997uy,Giovannini:1997eg}.
As understood more recently, it is a direct consequence of the second 
law of thermodynamics that the viscous relativistic hydrodynamics of 
any charged plasma with triangle anomaly carries currents with terms 
proportional to the vorticity and proportional to an external magnetic 
field~\cite{Son:2009tf}. Rather than being set by new material properties,
the strength of these anomalous contributions is given in terms of known 
thermodynamic functions and it is generally finite for non-vanishing
chemical potentials.  Anomalous hydrodynamics therefore must 
display currents with features that do not have a direct counterpart 
in plasmas without quantum anomalies. The charged plasmas of quantum 
field theories with gravity duals provide examples for which these features 
arise in explicit calculations~\cite{Erdmenger:2008rm,Banerjee:2008th,Landsteiner:2011cp,Landsteiner:2011iq,Neiman:2010zi}. There is also significant
progress in understanding how anomalous hydrodynamics arises in the long 
wavelength limit of classical kinetic theory supplemented by the notion of Berry
curvature~\cite{Son:2012wh,Stephanov:2012ki,Son:2012zy}, how it emerges 
in the quantum kinetic approach~\cite{Gao:2012ix,Chen:2012ca}, and which
terms arise in anomalous hydrodynamics beyond the first order in viscous 
corrections~\cite{Kharzeev:2011ds}.

In the quark gluon plasma of the phenomenologically realized quantum
field theory of QCD coupled to electrodynamics, the fermionic degrees 
of freedom carry both, the conserved electromagnetic and the anomalous 
axial charge. The collective close-to-equilibrium dynamics of the quark 
gluon plasma provides therefore a special case of anomalous 
hydrodynamics~\cite{Sadofyev:2010pr}:
For non-vanishing chemical potential of the axial charge, $\mu_A$, 
an external magnetic field $\vec{B}$ or a vorticity $\vec{\omega}$ 
in the fluid induces an electromagnetic vector current 
$\vec{j}_V$. And for non-vanishing 
vector chemical potential $\mu_V$, vorticity or an external magnetic
field induces an axial current  $\vec{j}_A$. 

The question to what extent these characteristic features of the chiral anomaly could lead to 
experimentally accessible signatures in the plasma produced in ultra-relativistic heavy ion collisions 
is currently at the focus of an intense theoretical debate~\cite{Kharzeev:2007tn,Kharzeev:2007jp,Fukushima:2008xe,Bzdak:2009fc,Bzdak:2010fd,Schlichting:2010qia,KerenZur:2010zw,Kharzeev:2010gd,Kharzeev:2010gr,Burnier:2011bf,Burnier:2012ae,Hongo:2013cqa}
and of experimental searches \cite{Voloshin:2004vk,Selyuzhenkov:2005xa,Abelev:2009ac,Abelev:2009ad,Mohanty:2011nm,Ke:2012qb,Adamczyk:2013gv,Adamczyk:2013kcb,Adamova:2012md}.  
Several proposed signatures of anomalous hydrodynamics are based on the idea that the
large $\vec{B}$-field in ultra-relativistic heavy ion collisions induces an axial charge asymmetry 
between the two hemispheres separated by the reaction plane. 
Originally, experimental searches focussed mainly on a scenario~\cite{Kharzeev:2007tn}
according to which these axial charges, due to interactions with non-trivial gauge configurations,
transferred into an electric charge asymmetry that would be visible event-by-event 
in an electric dipole moment. However, in this proposal the transfer of axial to electrical charge asymmetry involves physics that is not encoded in anomalous QCD hydrodynamics and that 
has not been elaborated on in a dynamical framework so far. Although the mechanism is
conceivable on general grounds, its efficiency in generating an electric charge difference is thus 
difficult to quantify. This is different for the mechanism of electric charge separation via the chiral 
magnetic wave (CMW), proposed more recently in Refs.~\cite{Burnier:2011bf,Burnier:2012ae}. 
First, the main experimental signature for the CMW
is an electric quadrupole moment~\cite{Burnier:2011bf}, rather than the electric dipole moment 
considered in Ref.~\cite{Kharzeev:2007tn}. Second, the CMW is an unavoidable consequence of 
QCD hydrodynamics coupled to QED that follows from general principles without any 
model-dependent assumption (this can be seen e.g. from our generalization of the CMW
to expanding fluids in section~\ref{sec2} below). Thus, there is a controlled model-independent
dynamical framework, namely hydrodynamics, within which one can discuss the efficiency of
the proposed mechanism of electric charge separation in heavy ion collisions.~\footnote{The 
caveat here is that the application of fluid dynamics is limited to sufficiently late times when 
gradients are small, while the effect of the CMW arises at sufficiently early times when the 
magnetic field is large. In the numerical studies of sections~\ref{sec3} and ~\ref{sec4}, we shall
typically work with initialization times $\tau_0 = 0.1$ fm/c. We regard the extrapolation of 
hydrodynamics to these early times as the best possible dynamical formulation of the CMW 
available at present. Given that the main conclusions
of sections~\ref{sec3} and ~\ref{sec4} depend only on the 
logarithmic order of magnitude of charge separation, there is an a posteriori argument 
that the inclusion of non-equilibrium effects at early times should not affect these conclusions 
significantly.}
The main purpose of the present work is to contribute to this discussion. 

Our work is organized as follows: in section~\ref{sec2}, we derive the equations of
motion of the chiral magnetic wave as a charge perturbation for the simplified case of 
an expanding Bjorken-type model. We explain how this leads to a transparent
model in which the efficiency of electric charge separation in response to an external magnetic 
field can be studied analytically and numerically.
After discussing in section~\ref{sec3a} an approximate analytical solution of the chiral magnetic wave 
that reveals already the main features of the full solution, we turn in sections~\ref{sec3b} and ~\ref{sec4}
to a numerical study of how efficient the chiral magnetic wave is in separating electric charges. 
We discuss how our results are related to (and consistent with) previous discussions of the
chiral magnetic wave, and we discuss conclusions. 

 \section{Anomalous hydrodynamics}
 \label{sec2}
 We consider a model of the energy-momentum tensor $T^{\mu\nu}$ of hot 
 QCD matter in an external electro-magnetic field $F^{\mu\nu}$. This matter carries both electric charge (written here in terms of the vector current 
 $j_V^{\mu} = \bar\psi \gamma^{\mu} \psi$) and axial charge (with axial 
 current  $j_A^{\mu} = \bar\psi \gamma^{\mu}\, \gamma^5 \psi$). The electro-magnetic field couples to the vector current,
and it determines the divergence of the anomalous axial current. 
The fluid dynamic equations of motion of this system read 
\begin{equation}
\begin{array}{lcl}
\nabla_{\mu}T^{\mu\nu} & = & Q_V F^{\nu\lambda} j_{V\lambda}\, ,\\
\nabla_{\mu} j_V^{\mu} & = & 0\, , \\
\nabla_{\mu} j_A^{\mu} & = & C E.B \, ,
\label{eq1}
\end{array}
\end{equation}
where the external electric and magnetic field is defined via a tensor decomposition of $F^{\mu\nu}$ with respect to the flow field $u^\mu$,
$E^{\mu}=F^{\mu\nu}u_{\nu}$ and $B^{\mu}=\frac{1}{2}\epsilon^{\mu\nu\alpha\beta}u^{\nu}F_{\alpha\beta}$. The coefficient $C$ denotes the strength of the abelian anomaly. In the following
$C = e/2\pi^2$ and we use for simplicity 
a vector charge $Q_V=e$. This neglects that different quark flavors carry
different fractional electric charge. (To avoid this simplification, one would have to 
introduce flavor-dependent
currents and chemical potentials.) It will be sometimes convenient to work in the
basis of left- and right-handed currents,  
\begin{equation}
 j_R^{\mu}=\frac{1}{2}\left(j_V^{\mu}+j_A^{\mu}\right) \quad j_L^{\mu}=\frac{1}{2}\left(j_V^{\mu}-j_A^{\mu}\right)\, .
 \label{eq3}
\end{equation}
Requiring that the second law of thermodynamics is satisfied locally, i.e. that the divergence of the entropy current is positive semi-definite everywhere, one
can write equations (\ref{eq1}) explicitly in the gradient expansion of viscous
fluid dynamics. Following Son and Surowka~\cite{Son:2009tf}, 
the energy momentum tensor and currents read to first order~\cite{Sadofyev:2010pr}  
\begin{eqnarray}
T^{\mu\nu} &=& (\epsilon+P)u^{\mu}u^{\nu}+Pg^{\mu\nu}+\tau^{\mu\nu}\, ,\\
j_{R,L}^{\mu} &=& n_{R,L} u^{\mu}+\nu_{R,L}^{\mu} \, ,
\label{eq4}
\end{eqnarray}
where
\begin{eqnarray}
\tau^{\mu\nu}&=&-\eta_s\Delta^{\mu\alpha}\Delta^{\nu\beta}\left(\nabla_{\alpha}u_{\beta}+\nabla_{\beta}
u_{\alpha}\right) \nonumber \\
&& -\left(\zeta-\frac{2}{3}\eta_s\right)\Delta^{\mu\nu}\nabla\cdot u 
\label{eq5}\, ,\\
\nu_{R,L}^{\mu}&=&-\frac{\sigma}{2}\left(T \Delta^{\mu\nu}\partial_{\nu}\left(\frac{\mu_{R,L}}{T}\right)+E^{\mu}\right) \nonumber \\
&& +\xi_{R,L}\omega^{\mu}+\xi_{R,L}^{(B)}B^{\mu}\, .
\label{eq6}
\end{eqnarray}
Here, the non-ideal part $\tau^{\mu\nu}$ of the energy-momentum tensor 
is characterized by the shear viscous transport coefficient $\eta_s$ and the 
bulk viscous coefficient $\zeta$. Up to first order, $\tau^{\mu\nu}$ can be 
written in terms of gradients of the flow field $u^\mu$ and the projector 
$\Delta^{\mu\nu}=g^{\mu\nu}+u^{\mu}u^{\nu}$. 
The viscous part $\nu_{R,L}^{\mu}$ of the left- and right-handed currents 
$j_{R,L}^{\mu}$ contains a conventional term that is proportional to the
vector charge conductivity $\sigma$. This contribution describes how the currents 
flow along an electric field or a gradient in the chemical potentials.  
What renders the hydrodynamics of the system (\ref{eq1}) anomalous is the 
second line of (\ref{eq6}). We focus on the term that induces currents parallel
to the magnetic field. Following Refs.~\cite{Son:2009tf,Sadofyev:2010pr}  but
keeping all terms, we find
\begin{eqnarray}
\xi_{R,L}^{(B)}&=&\frac{1}{2}C\left(\pm \mu_{R,L}-\frac{1}{2}\frac{n_{R,L}(\mu_R^2-\mu_L^2)}{\epsilon+P}\right)\nonumber\\
&+&\left(\pm Tf'\left(\textstyle{\frac{{\mu}_R-{\mu}_L}{T}}\right)
          -\frac{n_{R,L}T^2f\left(\textstyle{\frac{{\mu}_R-{\mu}_L}{T}}\right)}{\epsilon+P}\right)\, .
\label{eq7}
\end{eqnarray}
Here, the notation $\propto \pm \mu_{R,L}$ on the right hand side is a shorthand
denoting $+ \mu_R$ ($- \mu_L$) for $\xi_{R}^{(B)}$ ($\xi_{L}^{(B)}$). 
In this compact notation, $f$ is an additional integration constant that, as first pointed out 
in~\cite{Neiman:2011mj}, is not constrained by the thermodynamic
consistency relations for $\xi_{R,L}^{(B)}$~\cite{Sadofyev:2010pr}. 
The two terms proportional to $\textstyle{\frac{n_{R,L}}{\epsilon+P}}$ arise from 
rewriting terms proportional to $\partial T/\partial \mu_{R,L}$ with the help of
thermodynamic relations; we assume that these terms are much smaller than
the other two terms in (\ref{eq7}). More generally, we shall consider in the following situations with small charge densities and small 
chemical potentials, such that 
\begin{eqnarray}
   \nu_V^\mu &=& \left(\xi_{R}^{(B)} + \xi_{L}^{(B)} \right) B^\mu 
                            + O(\omega^\mu, \sigma) 
                            \simeq C\, \mu_A \, B^\mu\, ,
                            \nonumber \\
   \nu_A^\mu &=& \left(\xi_{R}^{(B)} - \xi_{L}^{(B)} \right) B^\mu 
                            + O(\omega^\mu) \nonumber \\
                      &\simeq& \left( C\, \mu_V \, +  
                      2\, T\, f'\left(\textstyle{\frac{{\mu}_A}{T}}\right) \right)  B^\mu\, .                
                      \label{eq8}  
\end{eqnarray}
Usually, discussions of experimental signatures neglect the unknown but 
thermodynamically allowed function $f\left(\textstyle{\frac{{\mu}_A}{T}}\right)$ 
in equations (\ref{eq7}) and (\ref{eq8}). Then, an external magnetic field induces an axial current
in the presence of a finite vector charge density. This phenomenon is referred to
as chiral magnetic effect (CME). Also, any time-dependent axial current will 
change the axial chemical potential $\mu_A$, and it will thus induce a vector 
current $\nu_V^\mu$. This is referred to as charge separation effect (CSE). 
The combination of both equations is the starting point for the discussion of 
the chiral magnetic wave (CMW). 

One may wonder to what extent for small chemical potential and charge densities
a non-vanishing function $f$ could change this picture. By construction, $\nu_V^\mu$
will remain unchanged, so charge separation along $B^\mu$ continues to take place 
for non-vanishing axial chemical potential. A negative term 
$\propto Tf'\left(\textstyle{\frac{{\mu}_A}{T}}\right)$, however, would counteract
the induction of the axial current, and a positive contribution would enhance it. 
We note that for the vortical effect, a corresponding term
was found to be non-vanishing in a holographic model~\cite{Landsteiner:2011iq}.
Here, we follow common practice and we neglect contributions proportional to $f$, but 
we argue that a better understanding of these terms in QCD is desirable.
 
\subsection{A Bjorken-type model}
For the strongly expanding system created in heavy ion collisions, the solution of
the equations of motion of viscous relativistic fluid dynamics is complicated, and
a fully realistic treatment requires numerical simulations. However, analytically
treatable models, and in particular Bjorken-type models, have contributed in the past
towards anticipating and understanding results of full fluid dynamical simulations. 
We expect that they can also play a useful role in understanding the dynamics of the
QCD fluid coupled to strong external electromagnetic fields, where numerical 
studies are at the very beginning~\cite{Hongo:2013cqa}. To this end, we formulate
and study here a simplified fluid dynamical model of the chiral magnetic wave 
that retains essential aspects of a strongly expanding fluid. 

In general, analytically treatable models of fluid dynamics assume additional symmetries 
in the initial conditions that are preserved during evolution and thus simplify the
equations of motion. What is then simplified is not the dynamical 
treatment of the QCD fluid, but the description of the initial conditions
under which this fluid is prepared. In practice, the symmetries assumed to this end 
in Bjorken-type models~\cite{Bjorken:1982qr} are longitudinal boost-invariance and
invariance under azimuthal rotations or under transverse translations. If all these
symmetries are invoked, one arrives at a 1+1 dimensional model that is analytically
solvable in ideal and first order viscous fluid dynamics. There is also a generalization of Bjorken-type models due to Gubser~\cite{Gubser:2010ze}, with analytical solutions of fluid dynamics that satisfy conformal symmetry. 

Formulating a simplified model of the chiral magnetic wave is complicated
since the phenomenon breaks all symmetries on which simplifications of fluid 
dynamics are typically based: a realistic magnetic field is not Bjorken boost-invariant
and the resulting fluid dynamic response breaks symmetry under rotations in the
transverse plane. To arrive at a model requires therefore to make
simplifying assumptions not only about the initial conditions but also about the
dynamical treatment of the QCD fluid. The main assumption in the following will
be that the axial and vector charge densities and the corresponding chemical potentials 
are parametrically small so that one can treat them as symmetry-breaking
perturbations that have a negligible effect on the energy density and pressure 
of the system. This assumption can be realized for a QCD fluid whose energy 
density is dominated by gluonic degrees of freedom, and for which net 
charge densities and associated chemical potentials are small. One can then
consider energy density and pressure as providing parametrically large background 
fields for which initial conditions and subsequent evolution satisfy the 
symmetries of Bjorken's model to leading order in $\mu/T$. In particular, to leading
order in $\mu/T$, the velocity $u^{\mu}$ that defines the 
rest frame of the energy density $\epsilon = u_{\mu}\, T^{\mu\nu}\, u_{\nu}$ 
takes at all space-time points the boost-invariant form
\begin{equation}
	u^{\mu} = \left( t/\tau,0,0,z/\tau\right) = \left(\cosh \eta, 0, 0, \sinh \eta\right)\, ,
\end{equation}
where $z$ is the beam direction, $\tau$ the proper time and $\eta$ the space-time
rapidity. Charges propagate as small perturbations (subleading in $\mu/T$)
on top of this Bjorken background field, and their propagation is seen in the dissipative parts 
$\nu_{R,L}^\mu$ of the currents as well as in the position 
dependence of the charge densities $n_{R,L}$ in (\ref{eq4}). It is only these latter terms
that break the symmetries of the Bjorken model in response to an external magnetic
field. 

For a constant external magnetic field $B_x$, we can write to leading order in $\mu/T$ 
eqs.~(\ref{eq1}) - (\ref{eq6}) as four equations for the energy density 
$\epsilon$ and pressure $P$, (we use $\bar{\mu}_{R,L} \equiv \mu_{R,L}/T$)
\begin{eqnarray}
	&&\partial_{\tau}\epsilon+\frac{\epsilon+P}{\tau}-
	\frac{\textstyle{\frac{4}{3}}\eta_s+\zeta}{\tau^2}
	= B_x\, \sigma T \sinh\eta \partial_y\bar{\mu}_V\, ,
	\label{eq10}\\
	&&\partial_x P = 0\, ,
	\label{eq11}\\
	&&\partial_y P =  B_x\, \sigma T \frac{1}{\tau} 
	\left( \cosh\eta\partial_{\eta}\bar{\mu}_V-n_V \sinh\eta\right)\, ,
	\label{eq12}\\
	&&\partial_\eta P = -  B_x\, \sigma T \tau \partial_y\bar{\mu}_V \, ,
	\label{eq13}
\end{eqnarray}
and two equations for the time evolution of the charge densities,
\begin{eqnarray}
	&& \partial_{\tau}n_{R,L}+\frac{n_{R,L}}{\tau} 
	- \sigma \vec{\nabla}\left(T\vec{\nabla}\bar{\mu}_{R,L}\right) 
	\nonumber\\
	&&\qquad \qquad \qquad + B_x\, \cosh\eta 
	\partial_x \xi_{R,L}^{(B)} = 0\, .
	\label{eq14}
\end{eqnarray}
These equations are not yet self-consistent since they were derived assuming
longitudinal boost-invariance for $u^{\mu}$, $P$ and $\epsilon$ while they
contain terms with explicit $\eta$-dependence that break 
longitudinal boost-invariance. The reason for this is, of course, that 
(\ref{eq10})-(\ref{eq14}) were derived for 
a constant magnetic field $B_x$ that breaks longitudinal boost-invariance. 
To obtain a consistent, longitudinally boost-invariant set of differential 
equations, we could choose either a magnetic field of longitudinally boost-invariant 
form that coincides at $\eta = 0$ with the physical value of $B_x$ at mid-rapidity. 
This would give rise to a consistent dynamics that is physically meaningful at 
$\eta = 0$. Equivalently, we can simply limit our discussion of equations 
(\ref{eq10})-(\ref{eq14})  to the transverse plane at space-time rapidity 
$\eta = 0$, for which all terms on the right hand side of 
equations (\ref{eq10})-(\ref{eq13}) vanish. For the terms $\propto \partial_y\bar{\mu}_V$
and $\propto \partial_{\eta}\bar{\mu}_V$, this follows from the fact that even
in the presence of a constant magnetic field, the system is symmetric under translation 
in the transverse direction orthogonal to the magnetic field and under 
reflection $\eta \to - \eta$.
For $\eta = 0$, we therefore find for energy
density and pressure the equations of motion of the Bjorken model to first
order in viscous hydrodynamics,
\begin{eqnarray}
	&&\partial_{\tau}\epsilon+\frac{\epsilon+P}{\tau}
	-\frac{\textstyle{\frac{4}{3}}\eta_s+\zeta}{\tau^2}
	= 0\, ,
	\label{eq15}\\
	&&\vec\partial P = 0\, ,
	\label{eq16}
\end{eqnarray}
supplemented by a simple time evolution for the charge densities,
\begin{eqnarray}
	 \partial_{\tau}n_{R,L}+\frac{n_{R,L}}{\tau} 
	- \sigma \nabla^2 \mu_{R,L} 
	 \pm B_x\, 
	\partial_x \xi_{R,L}^{(B)} = 0\, .
	\label{eq17}
\end{eqnarray}
Equations (\ref{eq15})-(\ref{eq17}) are the starting point
of the following discussion. We recall tht they provide a simplified consistent 
dynamical treatment if restricted to mid-rapidity. The following
analytical and numerical studies will be limited to the case where the energy density
is translationally invariant in the transverse $(x,y)$-plane and where the charge distributions
are invariant in the transverse direction $y$ orthogonal to the direction of the magnetic field. 
The advantage of translation invariance with respect to $y$ is that we arrive at a very 
transparent model for the efficiency of electric charge separation via the CMW. 
%
\subsection{The chiral magnetic wave in an expanding fluid}
Equation (\ref{eq17}) describes the hydrodynamic propagation of chiral charge
density in a Bjorken-expanding background. To make contact with earlier discussions
of the chiral magnetic wave, we work in the limit of small chemical potential when
\begin{eqnarray}
	\mu_{R,L} \simeq 2\, \alpha\, n_{R,L}\, ,
	\qquad \alpha \equiv \frac{\partial \mu_{R,L}}{2\partial n_{R,L}}
	\Bigg\vert_{n_{R,L}=0}\, .
	\label{eq18}
\end{eqnarray}
Keeping only the dominant first term of equation (\ref{eq7}), $\xi_{R,L}^{(B)} = 
\pm \textstyle\frac{1}{2} C \mu_{R,L}$, we obtain 
\begin{equation}
    \partial_{\mu} \left(n_{R,L} u^{\mu}\right) \pm B_x C \alpha \partial_x n_{R,L}
    - 2 \sigma \alpha \Delta n_{R,L} = 0.
    \label{eq19}
\end{equation}
According to this equation of motion, right-handed (left-handed) charge density is 
transported with velocity 
\begin{equation}
	v_x = B_x\, C\, \alpha
	\label{eq20}
\end{equation}
along (opposite to) the magnetic field. This charge transport is accompanied by
diffusion characterized by the diffusion constant
\begin{equation}
 D = 2\, \sigma\, \alpha\, .
 \label{eq21}
\end{equation}
We note that equation (\ref{eq19}) derived here is a natural generalization 
of the equation for a chiral magnetic wave written in 
Ref.~\cite{Burnier:2011bf,Burnier:2012ae} for the case of a static medium. 
Indeed, inserting the velocity profile for a time-independent medium,
$u^{\mu} = (1,0,0,0)$, into (\ref{eq19}) one finds the form discussed in 
Ref.~\cite{Burnier:2011bf,Burnier:2012ae}, 
\begin{equation}
 \left(\partial_t \pm v_x \partial_x - D \partial_x^2\right) n_{R,L} = 0\, .
 \quad \hbox{(static medium)}
 \label{eq22}
\end{equation} 
The main features of charge transport in a static medium can be 
inferred from general considerations without explicit calculation. In particular, one 
knows that the velocities $u^\mu$, $u^\mu_{\rm (E)}$ of the fluid dynamic frames 
comoving with energy density (Landau frame) and with a charge density 
(Eckhart frame) are related in first order hydrodynamics by the linear
relation (see e.g. Ref.~\cite{Muronga:2003ta}) 
\begin{equation}
u^\mu_{{R,L}({\rm E})} = u^\mu + \frac{\nu^{\mu}_{R,L}}{n_{R,L}}\, .
\label{eq23}
\end{equation}
The velocities of charge transport are then given by the difference between the
velocity that defines the Landau frame of comoving energy energy density
and the Eckhart frames of comoving left (+) or right (-) handed charge densities.
Inserting the above relations, one checks easily the expectation that the velocity
of charge transport 
\begin{equation}
	\pm v_x = u^\mu_{{R,L}({\rm E})} - u^\mu = \frac{1}{2} C \frac{\mu_{R,L}}{n_{R,L}} B_x
	= B_x C \alpha
	\label{eq24}
\end{equation}
is determined by the transverse boost from the Landau
frame of comoving energy density to the Eckhart frame that
comoves with left (+) or right (-) handed charges. The velocity $v_x$
of charge transport derived for the Bjorken-type model in (\ref{eq20})
equals that obtained from this general consideration.
In Refs.~\cite{Burnier:2011bf,Burnier:2012ae}, also the presence of a diffusion 
term $\propto \Delta n_{R,L}$ was argued for on physical grounds.
In summary, we have shown in this subsection that the fluid dynamic equations
of motion (\ref{eq15})-(\ref{eq17}) are a natural generalization of the chiral magnetic 
wave equation of Ref.~\cite{Burnier:2011bf,Burnier:2012ae} to the case of a 
Bjorken-expanding background field, and we have recalled how the physics encoded
in these equations is related to general considerations about charge transport. 

\section{Solutions to the CMW in an expanding background}
\label{sec3}

In general, solutions to the fluid dynamic equations of motion depend on the initial
conditions for all fluid dynamic fields, the equation of state and the value of transport coefficients, and the strength of external fields. As for the equation of state, there are lattice 
simulations~\cite{Allton:2005gk} that determine for two-flavor QCD up to sixth order 
($k=3$) the coefficients $C_k$ in the power series
\begin{equation}
	P(T,\mu_R,\mu_L) = T^4 \sum_{k} C_k \left[ 
	   \left(\frac{\mu_R}{T}\right)^{2k} + \left(\frac{\mu_L}{T}\right)^{2k}\right]\, .
	   \label{eq25}
\end{equation}  
As we shall work in the present case for small chemical potentials only, we shall 
focus on the first two terms in this series, setting $C_k =0$ for $k\geq 2$. This is
also justified by the observation in Ref.~\cite{Allton:2005gk} that the coefficients
$C_k$ of higher orders $k\geq 1$ are numerically smaller by factors $O(10^{-k})$. 
Numerically, it was found that at temperatures well above $T_c$, the coefficients
$C_0$ and $C_1$ are approximately 80 \% of the value of a free gas of gluons and
$n_f=2$ quarks~\cite{Allton:2005gk}. This is in line with results for field theories 
with gravity duals for which the equation of state is known to vary between the limits 
of vanishing and infinite coupling by a constant factor $3/4$~\cite{Gubser:1996de} or 
by a factor close to $3/4$~\cite{Nishioka:2007zz}. 
For numerical studies, we shall therefore choose (\ref{eq25}) with
\begin{eqnarray}
	C_0 &=& 0.8 \left( \frac{8\pi^2}{45} + \frac{\pi^2}{15} \frac{7}{4} n_f\right)\, ,
	\label{eq25a}\\
	C_1 &=& 0.8 \frac{1}{4} n_f\, .
	\label{eq25b}
\end{eqnarray}
Furthermore, to allow for analytical calculations, 
we consider in the following an ideal equation of state $\epsilon = 3\, P$.

With the help of the thermodynamic relation, $dP = s\, dT + n_R d\mu_R + n_L d\mu_L$
and the equation of state (\ref{eq25}), we can write the equations of motion (\ref{eq15})-(\ref{eq17}) as coupled partial differential equations for the temperature and the chemical
potentials $\mu_R$ and $\mu_L$.  Using ${\cal M}^2 \equiv \mu_R^2+\mu_L^2$, we write
\begin{eqnarray}
 &&\left(2 C_0 T^3+C_1T{\cal M}^2\right)\partial_{\tau}T
  + \frac{1}{2} C_1T^2\partial_{\tau} {\cal M}^2  
   \label{eq26} \\ 
  &&\quad + 
  \frac{2}{3\tau}\left(C_0T^4+C_1T^2{\cal M}^2 \right)
 =\frac{4 \textstyle{\frac{\eta}{s}}}{9\tau^2}  \left(2 C_0 T^3+C_1T{\cal M}^2 \right)\, ,
\nonumber 
\end{eqnarray}
and
\begin{eqnarray}
&& \partial_{\tau}\mu_{R,L} +  \frac{\mu_{R,L}}{\tau} + 2 \frac{\mu_{R,L}}{T} \partial_{\tau} T 
\pm \frac{1}{4} \frac{C}{C_1} \frac{B_x}{T^2} \partial_x\mu_{R,L}
\nonumber \\
&& \qquad  - \frac{\sigma}{2C_1T^2} \partial_x^2\mu_{R,L} = 0\, .
\label{eq27}
\end{eqnarray}
These equations of motion can be solved numerically for arbitrary initial conditions, and
we shall discuss such numerical results in the following. First, however, we turn to an
approximate analytical solution that will turn out to illustrate the main features of the full 
numerical result.

\subsection{Approximate analytical solution for Gaussian initial condition}
\label{sec3a}
In the limit of small charge densities and negligible viscosity, the solution to the equation of
motion (\ref{eq26}) is the Bjorken solution up to parametrically small corrections,
\begin{equation}
T = T_0\, \left(\frac{\tau_0}{\tau}\right)^{1/3} + O\left( \textstyle\frac{\mu_{R,L}}{T}, 
\textstyle{\frac{\eta}{s}}\right) \, .
\label{eq28}
\end{equation}
Entering with this expression equation (\ref{eq27}) and neglecting all non-linear,
higher order terms in the charge density, one finds
\begin{equation}
\partial_{\tau} \tilde{\mu}_{R,L}\pm \frac{C}{4C_1}\frac{\tau^{2/3}\, B_x}{T_0^2\tau_0^{2/3}} 
\partial_x  \tilde{\mu}_{R,L} 
- \frac{\tau^{2/3}\sigma}{2C_1 T_0^2\tau_0^{2/3}}\partial_x^2 \tilde{\mu}_{R,L}
=0 \, .
\label{eq29}
\end{equation}
Here, we have separated the leading time dependence of the charge
densities, that is due to longitudinal Bjorken expansion, 
\begin{equation}
  {\mu}_{R,L}(\tau,x)  = \tilde{\mu}_{R,L}(\tau,x)\, \left(\frac{\tau_0}{\tau}\right)^{1/3}\, .
  \label{eq30}
\end{equation}
We consider first initial conditions of heavy ion collisions that show event-by-event
an initially negligible density $n_A$ of axial charges but a non-vanishing density 
of electric charges of approximately Gaussian shape, centered at initial time $\tau_0$ 
at transverse position $x = 0$. We want to understand how this electric charge
distribution evolves due to the effects of the QCD anomaly 
in the presence of a large and time-dependent magnetic field 
$B_x(\tau)$. The fluid dynamic problem studied here is simpler than the situation 
in heavy ion collisions in that it ignores transverse gradients in energy density and
pressure and thus retains a one-dimensional translational symmetry in the transverse
direction $y$. However, the problem is sufficiently complex to study the effects of 
charge separation along the magnetic field in a Bjorken expanding fluid. 
Interestingly, while the discussion of (\ref{eq29}) for general initial conditions requires 
numerical techniques, the solution for a Gaussian initial density distribution can be
given analytically
\begin{equation}
\mu_{R,L}(\tau,x)= \left(\frac{\tau_0}{\tau} \right)^{1/3} 
\frac{{\rm const.}}{R(\tau)} 
\exp\left[- \frac{1}{2}\frac{(x + s_{\pm}(\tau))^2}{R^2(\tau)}\right]\, .
\label{eq31}
\end{equation}
Here, the peak of the Gaussian distribution for right- and left-handed
charges evolves in time according to  
\begin{equation}
 s_{\pm}(\tau)
 =\mp \frac{1}{4}\frac{C}{C_1}\frac{\tau_0^{-2/3}}{T_0^2}
    \int_{\tau_0}^{\tau}\tau'^{2/3}B_x(\tau')d\tau'\, .
    \label{eq32}
\end{equation}
For the equation of state (\ref{eq25}), one finds $n_{R,L} = 2 C_1 T^2 \mu_{R,L}$
and therefore $\alpha = \textstyle{ \frac{\partial \mu_{R,L}}{2\partial n_{R,L}}}
= 1/ 4 C_1 T^2$. From this, one checks easily that the position of the peak moves
with the expected velocity (\ref{eq20}),
\begin{equation}
	\frac{ds_{\pm}(\tau)}{d\tau} = v_x(\tau) = B_x(\tau)\, C\, \alpha\, .
	\label{eq33}
\end{equation}
The spatial separation $s_+ - s_-$ of right- and left-handed charges is of direct 
phenomenological relevance. In the fluid dynamic model studied here, this separation
is proportional to a particular time-average of the time-dependent external magnetic 
field,  $\int_{\tau_0}^{\tau}\tau'^{2/3}B_x(\tau')d\tau'$. One may wonder whether this
finding is valid also outside the model-dependent framework in which we have derived
it here. In this context, we note that for small charge densities, one generally expects
a linear relation between chemical potentials and charges. In the absence of other
dimensionful parameters, this dependence must be of the form $n_{R,L} \propto
T^2 \mu_{R,L}$, and therefore $\alpha \propto 1/T^2$. For an expanding fluid, it is
then the time-dependence of the dimensionless ratio $B_x(\tau) / T^2(\tau)$ that 
determines the time-dependence of the velocity $v_x(\tau)$ with which right- and left-
handed charges separate. The time-dependence of the integrand of (\ref{eq32}) is
therefore generally expected for fluids with Bjorken expansion. 
We note as an aside that for the case of a fluid with transverse expansion, when one
expects $T(\tau) = T_0 \left(\tau_0/\tau\right)^\delta$, $1/3 < \delta < 1$, these 
considerations suggest that the spatial separation $s_+ - s_-$ will depend on a 
weighted time-average $\int_{\tau_0}^{\tau}\tau'^{2\delta}B_x(\tau')d\tau'$ of the
magnetic field. 

The square of the Gaussian width in (\ref{eq31}) broadens according to the differential
equation $\partial_\tau \left(\textstyle\frac{1}{R^2}\right) = - 
\textstyle\frac{\sigma}{C_1\, T^2} \left(\textstyle\frac{1}{R^2}\right)^2$.
In the present paper, we consider mainly the case of a temperature-independent electrical
conductivity, for which one finds
\begin{eqnarray}
  R^2(\tau)\Bigg\vert_{\sigma = const.} = R^2(\tau_0) +
  \frac{3}{5}\frac{\sigma}{C_1 T_0^2 \tau_0^{2/3} }
   (\tau^{5/3}-\tau_0^{5/3}) \, .
   \label{eq34}
\end{eqnarray}
We note, however, that in calculations of the conductivity $\sigma$ 
of conserved U(1) charges in theories with gravity dual, one finds a linear dependence 
on temperature, $\sigma = \sigma_0\, T$~\cite{CaronHuot:2006te,Mateos:2007yp},
and this is also expected on dimensional grounds. Including in the calculation of
$R^2(\tau)$ the time-dependence of $\sigma$ that would result from a
linear temperature dependence, one finds instead of (\ref{eq34}) a slightly weaker 
time-dependence
\begin{eqnarray}
  R^2(\tau)\Bigg\vert_{\sigma = \sigma_0\, T} = R^2(\tau_0) +
  \frac{3}{4}\frac{\sigma_0}{C_1 T_0 \tau_0^{1/3} }
   (\tau^{4/3}-\tau_0^{4/3}) \, .
   \label{eq35}
\end{eqnarray}
This illustrates that details of the diffusion will depend on the temperature-dependence
of the conductivity. However, these effects turn out to be numerically small, and we 
therefore fix in the following the conductivity to a value $\sigma = e^2\, 200$ MeV that
is consistent with the recent (quenched) lattice QCD studies~\cite{Ding:2010ga}, 
$\frac{1}{3} e^2\, T_c < \sigma < e^2\, T_c$, where 
$e^2 = 4\pi/137$. 

\subsection{Numerical Results for initial Gaussian vector charge distribution}
\label{sec3b}
We start by summarizing the input used in the following numerical study of the 
fluid dynamic equations of motion 
(\ref{eq15})-(\ref{eq17}) of the chiral magnetic wave. We use the ideal equation of
state $\epsilon = 3 P$ with pressure given in equation
(\ref{eq25}) and with non-vanishing coefficients (\ref{eq25a}), (\ref{eq25b}). The 
default values for the ratio of shear viscosity over entropy and for the electrical
conductivity are $\eta/s = 1/4\pi$ and  $\sigma = e^2\, 200$ MeV, respectively.
As default, we initialize the system with a temperature 
$T(\tau_0 = 0.1\, {\rm fm/c}) = T_0 = 650$ MeV that 
corresponds for a fluid with Bjorken expansion
to the initial temperature $T(\tau_0 = 0.6\, {\rm fm/c}) = 360$ 
MeV consistent with fluid dynamic simulations of heavy ion collisions at RHIC energy
($\sqrt{s_{\rm NN}} = 200$ GeV). 
We initialize the chemical potentials with a Gaussian profile (\ref{eq31}) of 
transverse width $R = 3$ fm. We use an 
initial strength $\mu_{R,L}(\tau_0,x=0) = 10$ MeV consistent with the assumption
that the contribution of charges to the pressure of the system is perturbatively
small, $\mu/T \ll 1$. We note, however, that the dependence on $\mu_{R,L}$ is 
linear for sufficiently small charge densities, and most of our results will be 
normalized such that they are independent of $\mu_{R,L}(\tau_0,x=0)$.
For the strength of the abelian anomaly, we use $C = \textstyle{\frac{e}{2\pi^2}}$; this means,
that we do not take the fractional charge of quarks into account. 

One aim of the following study is to arrive at some qualitative statement about how
phenomenologically testable effects of the chiral magnetic wave depend on the center 
of mass energy. To allow for this discussion, we include in the calculation the
$\sqrt{s_{\rm NN}}$-dependence of the initial temperature by requiring that the entropy 
of the system matches the $\sqrt{s_{\rm NN}}$-dependence of the measured charged 
particle multiplicity at mid rapidity,
$T^3(\tau = {\rm fixed}) = {\rm const.} dN_{\rm ch}/d\eta$. For the event multiplicities in 
heavy ion collisions published in Ref.~\cite{Back:2002wb,Aamodt:2010pb}, this implies that 
in comparison to the upper RHIC energy (Au+Au at $\sqrt{s_{\rm NN}} = 200$ GeV), the initial
temperature $T_0$ at fixed $\tau_0$ 
is a factor $\sim 1.35$ larger at LHC (Pb+Pb at $\sqrt{s_{\rm NN}} = 2.76$ TeV) 
and a factor $\sim 0.81$ smaller at lower RHIC or upper SPS energy of $\sqrt{s_{\rm NN}} = 17$ GeV.
Moreover, we choose for the external magnetic field in our calculations the dependence 
on center of mass energy $\sqrt{s_{\rm NN}}$, proper time $\tau$ and impact parameter $b$
expected for realistic heavy ion collisions. We calculate the strength of this magnetic field 
from the distributions of electric charges of spectators and participants 
according to the classical electrodynamic calculation described in the appendix of 
Ref.~\cite{Kharzeev:2007jp}, see Fig.~\ref{fig1}. 
%
\begin{figure}[h]
\begin{center}
\includegraphics[width=8.cm]{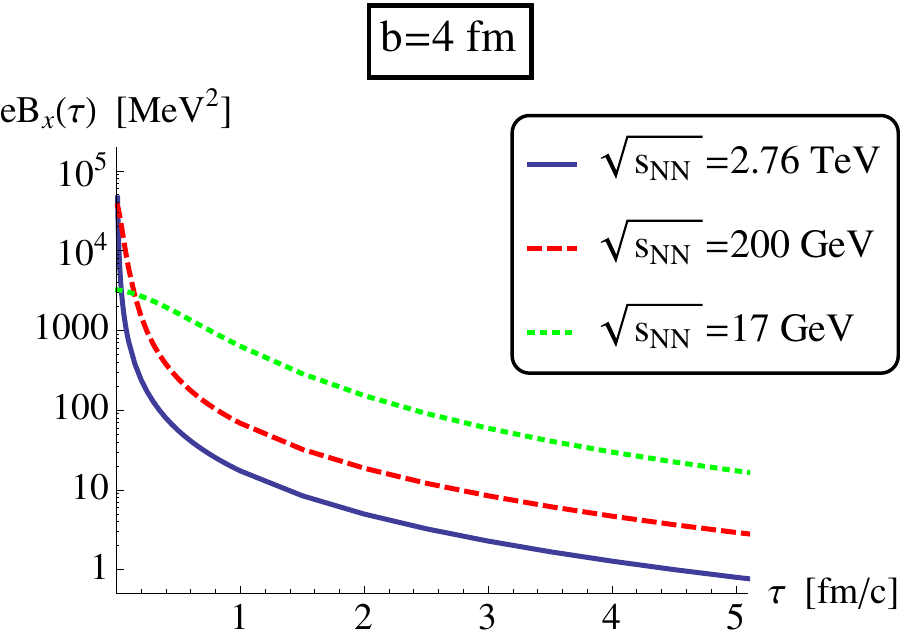}
\includegraphics[width=8.cm]{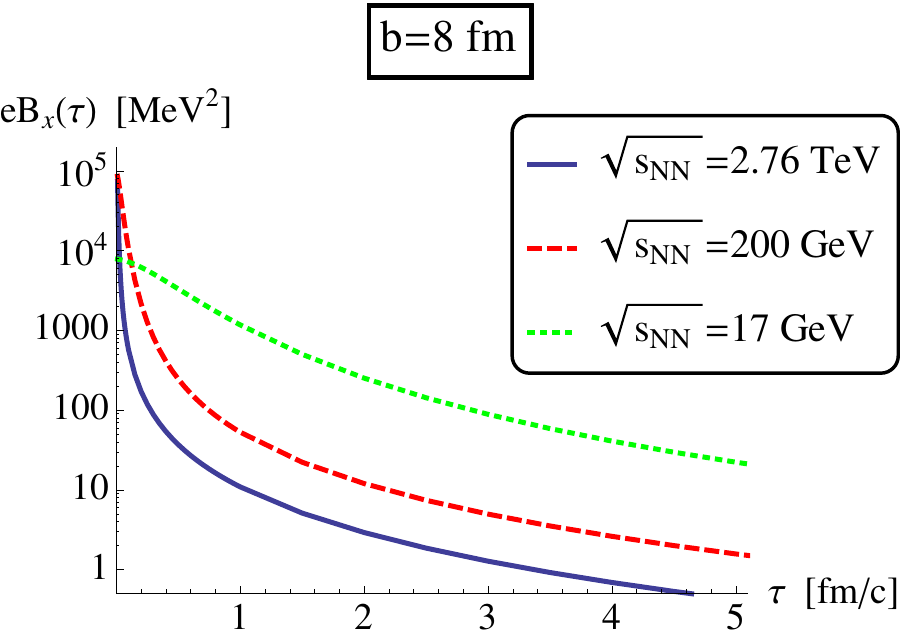}
\end{center}
\vspace{-0.5cm}
\caption{Time-dependent strength of the transverse component of the
external magnetic field obtained in a classical electromagnetic calculation~\cite{Kharzeev:2007jp} field for nucleus-nucleus collisions at different center of mass energies 
and for different impact parameter $b$.
}\label{fig1}
\end{figure}

The $\sqrt{s_{\rm NN}}$-dependence of the maximal value of the magnetic field 
strength and the temporal fall-off shown in Fig.~\ref{fig1} is in rough quantitative
agreement with numerical results of a microscopic transport 
model~\cite{Skokov:2009qp} for fixed target and RHIC energies, and with
calculations for RHIC and LHC energies~\cite{Mo:2013qya}. In comparison to 
a model of hadron string dynamics~\cite{Voronyuk:2011jd}, Fig.~\ref{fig1} overestimates 
the magnetic field strength somewhat. On the other hand, it was demonstrated in model
studies that event-wise fluctuations in the incoming 
charge distributions can lead to larger values of the magnetic field strength in 
some events~\cite{Bzdak:2011yy,Deng:2012pc}, and that they can also lead to very large 
fluctuations in the electric field strength. In summary, these findings suggest that the 
time-dependent magnetic field strengths plotted in Fig.~\ref{fig1} provide reasonable
ball-park estimates for the following discussion. One may wonder whether this 
conclusion could change significantly due to 
effects of magnetization of the QGP that may prolong the lifetime of the magnetic 
field~\cite{Tuchin:2013ie}. For this to happen, one would have to require a dynamical 
mechanism able to trap magnetic fields of strength $> 10^{17}$ Gauss in a fluid of 
quarks and gluons. A recent analysis suggests that at least for the values of the electric 
conductivity obtained from lattice QCD calculations, such effects of magnetization are 
relatively small~\cite{McLerran:2013hla}. In the following numerical studies, we therefore use
the strength and time-dependence of the $B_x$-field in Fig.~\ref{fig1}. Since the dependence
on impact parameter is relatively mild, we show results only for a default value of the $B_x$-field 
at  $b = 4$ fm. In the present model,
this magnetic field is position-independent and permeates the fluid at mid rapidity. 
%
\begin{figure}[h]
\begin{center}
\includegraphics[width=8.cm]{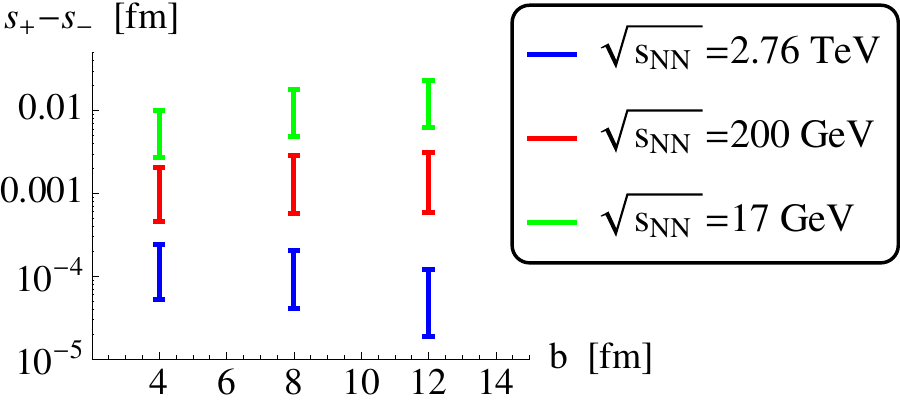}
\end{center}
\vspace{-0.5cm}
\caption{Spatial separation of right- and left-handed axial charge, occuring according to 
eq.~(\ref{eq32}) in the magnetic field of Fig.~\ref{fig1} between times $\tau_0$ and $\tau = 10$ fm/c.
The upper (lower) end of the uncertainty band corresponds to initial value $\tau_0 =$ 
$0.01$ fm/c ($0.1 fm/c$).
}\label{fig2}
\end{figure}

Fig.~\ref{fig1} shows clearly that while the peak strength of the magnetic field increases with 
increasing center of mass energy, its temporal fall-off is much faster for larger $\sqrt{s_{\rm NN}}$.
As a consequence, the external magnetic field at upper CERN SPS/ lower RHIC energy of 
$\sqrt{s_{\rm NN}} = 17$ GeV exceeds from times $\tau > 0.1 fm/c$ onwards the corresponding
external fields attained at upper RHIC and LHC energies.
Whether phenomenologically testable effects of the chiral magnetic wave are
maximized at higher or lower center of mass energy thus becomes a dynamical question. 
To inform us about his point, we plot in Fig.~\ref{fig2} the spatial separation $s_+ - s_-$ of 
right- and left-handed charges, evaluated according to equation (\ref{eq32}) for different 
centralities and different center of mass energies. Before commenting on the absolute size
of the spatial separation, let us comment on the observed dependencies. In particular, we find
that the spatial separation is larger at smaller center of mass energy. We recall that the velocity
of charge separation is in (\ref{eq32}) is $v_x(\tau) \propto B_x(\tau)/T^2(\tau)$. This explains
that the $\sqrt{s_{\rm NN}}$-dependence of $s_+ - s_-$ seen in Fig.~\ref{fig2} is mainly due to
the weaker temporal fall-off of the magnetic field at small $\sqrt{s_{\rm NN}}$, but the fact
that the temperature at fixed time $\tau$ decreases with decreasing $\sqrt{s_{\rm NN}}$ does
also play a role. Since $B_x(\tau)$ peaks at very small times, there is a significant 
$\tau_0$-dependence of $s_+ - s_-$. 
Initializing a hydrodynamic calculation at too short times $\tau_0$ is certainly questionable.
However, to quantify the uncertainty in varying $\tau_0$, it is useful to note that the integral
$\int_{\tau_0}^{\tau} \tau'^{2/3} B_x(\tau') d\tau'$ in (\ref{eq32}) stays finite even in the limit
$\tau_0 \to 0$. For this reason, we plot in Fig.~\ref{fig2} upper values that correspond to 
maximizing the contribution of this integral by choosing a very small lower integration boundary 
$\tau_0 = 0.01$ fm/c. Even with this optimistic ansatz, the spatial separation of right- and
left-handed charges during the evolution is only of order $O(10^{-2}\, {\rm fm})$ at SPS
energies and it is one (two) orders smaller at RHIC (LHC) energies. This raises the question
under which circumstances phenomenological consequences of the chiral magnetic wave
may become testable. 

\begin{figure}[h]
\begin{center}
\includegraphics[width=8.cm]{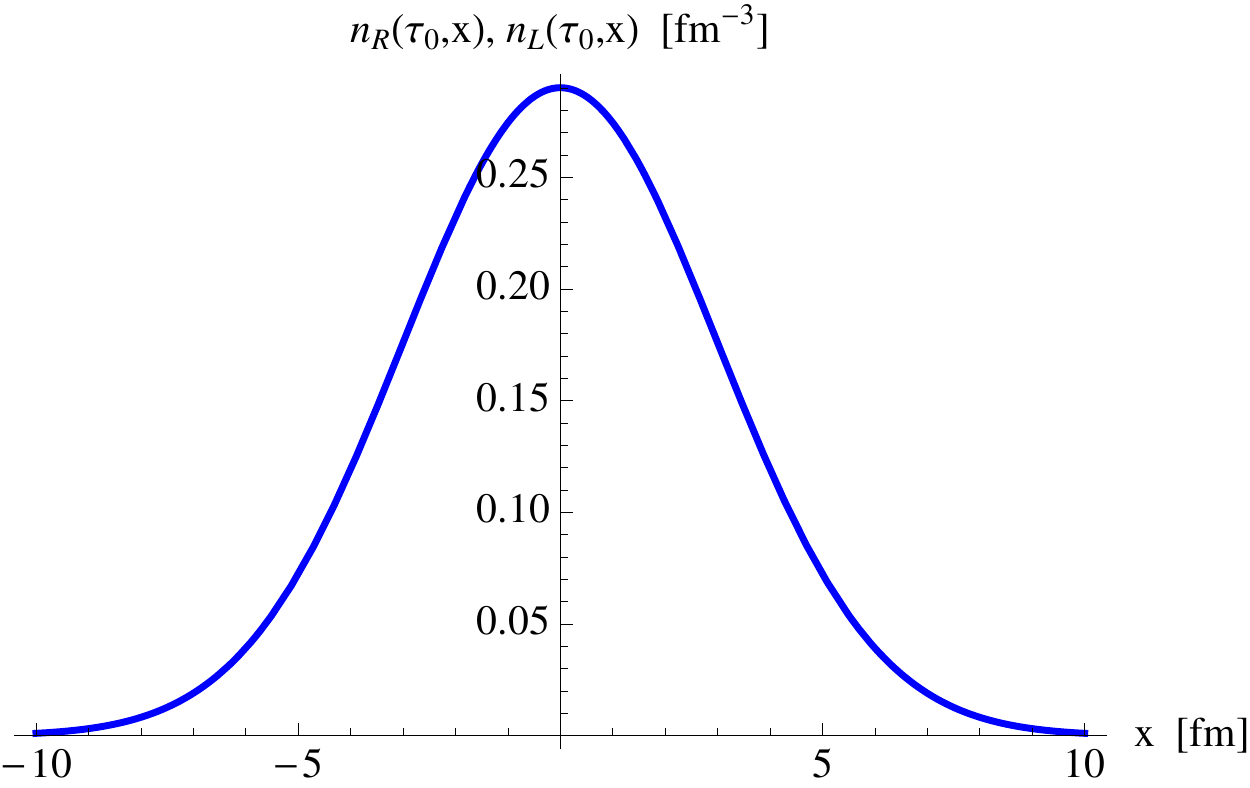}
\end{center}
\vspace{-0.5cm}
\caption{Initial condition for the distribution of charge densities $n_R$, $n_L$, 
corresponding to Gaussian distribution (\ref{eq31}) with norm
$\mu_{R,L}(\tau_0,x= 0.) = 10$ MeV and Gaussian width $R(\tau_0) = 3$ fm.
}\label{fig3}
\end{figure}
\begin{figure}[h]
\begin{center}
\includegraphics[width=8.cm]{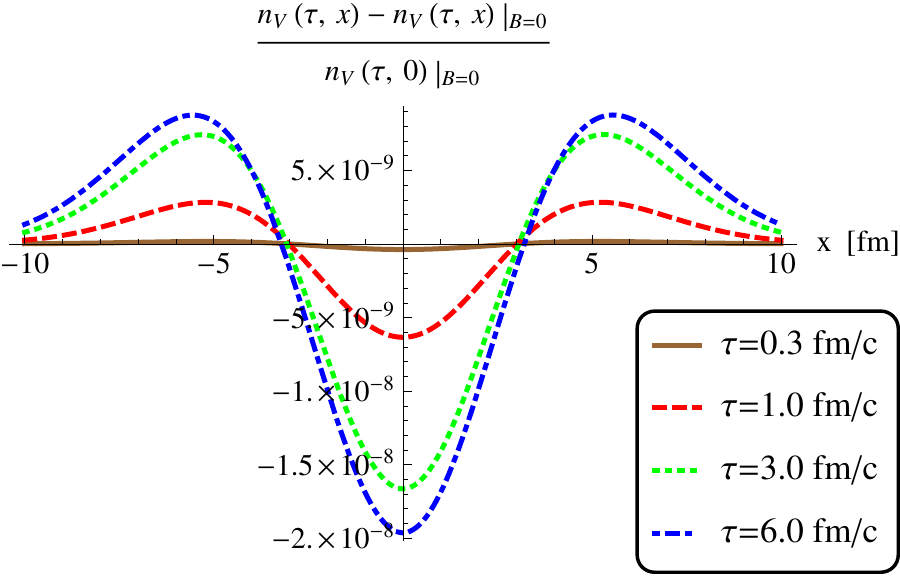}\\
\vspace{.5cm}
\includegraphics[width=8.cm]{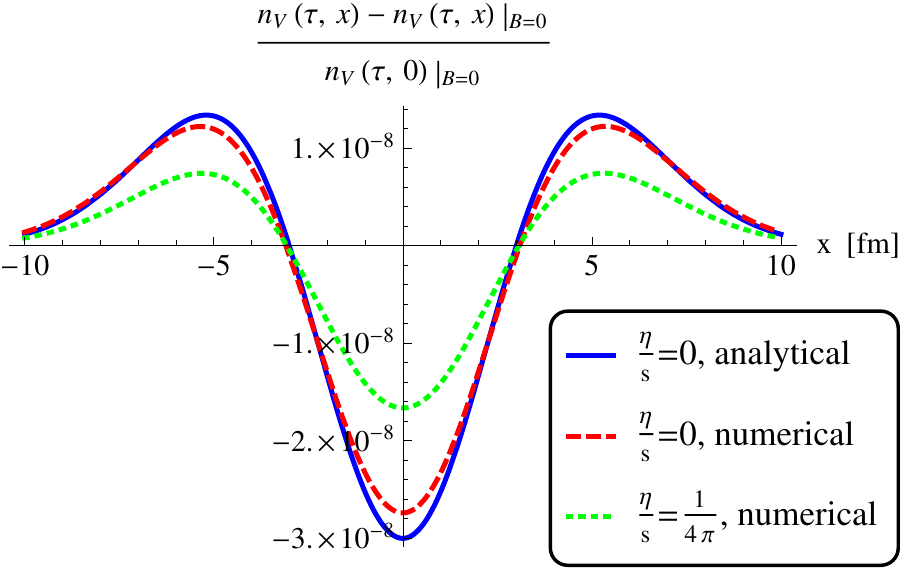}\\
\vspace{.5cm}
\includegraphics[width=8.cm]{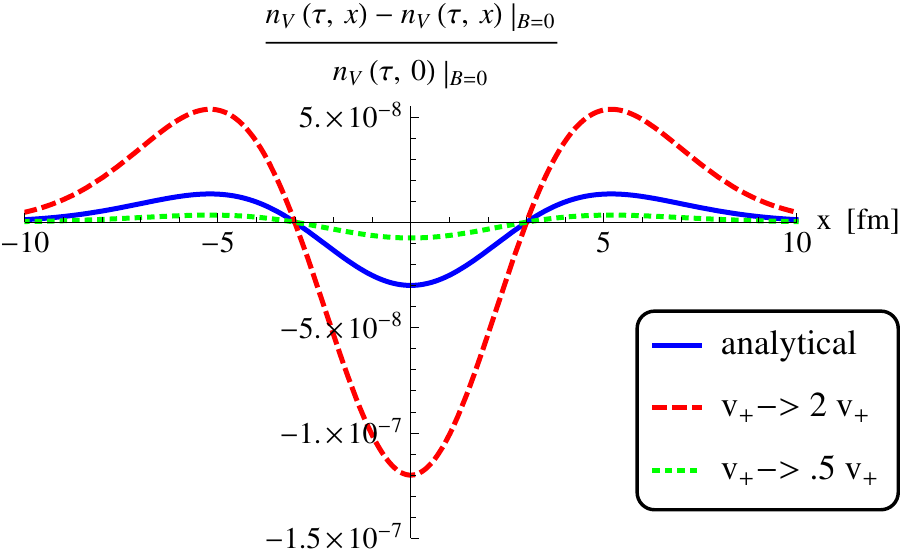}
\end{center}
\caption{The normalized and subtracted vector charge distribution for initial conditions corresponding 
to $b=4$ fm at $\sqrt{s_{\rm NN}} = 17$ GeV and initial charge distributions shown in Fig.~\ref{fig3}. 
Top panel: numerical solution of (\ref{eq26}),(\ref{eq27}) for different times into the evolution.
Middle panel: comparison of the analytical result (\ref{eq32})-(\ref{eq34}) with the full numerical
solution at vanishing and at finite shear viscosity. Lower panel: dependence of the analytical
result on the analytically determined transverse velocity of charge transport.
}\label{fig4}
\end{figure}
\begin{figure}[h]
\begin{center}
\vspace{.5cm}
\includegraphics[width=8.cm]{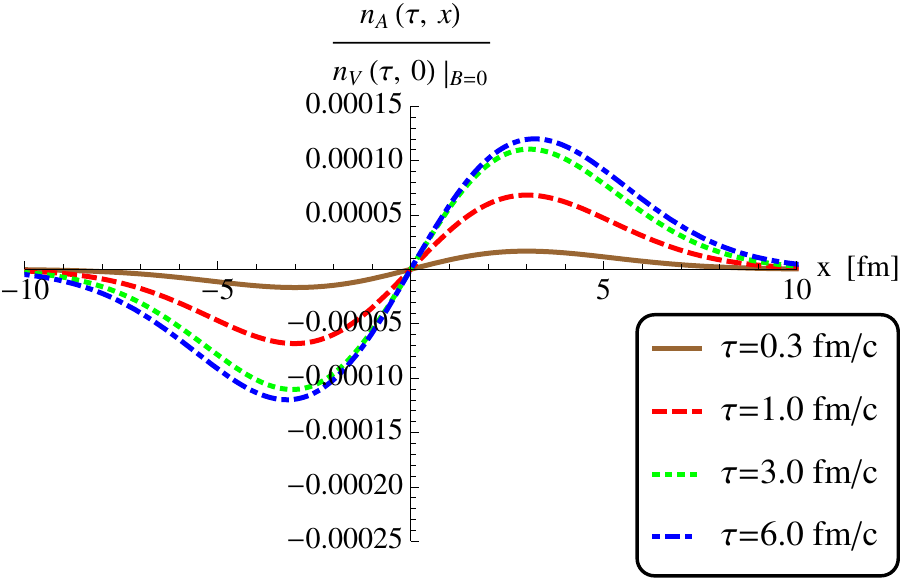}
\end{center}
\caption{The normalized axial charge distribution determined for the same input as the
top panel of Fig.~\ref{fig4}.
}\label{fig5}
\end{figure}

%
We now turn to the discussion of numerical solutions of the equations of motion (\ref{eq26}),
(\ref{eq27}) for the CMW. We first consider an initial condition that corresponds along the direction $x$ of
the magnetic field to a Gaussian distribution of vector charge $n_V = n_R + n_L$ and a 
vanishing distribution of axial charge $n_A = n_R - n_L$. Fig.~\ref{fig3} shows this distribution
for the default values of the thermodynamic variables discussed above. One sees that the default
values chosen here correspond to a small charge density, consistent with the idea that we can treat 
charge transport as a perturbation on top of a dynamically evolving background. According to 
Fig.~\ref{fig2} and the analytical considerations of section~\ref{sec3a}, we expect that the peak 
of the Gaussian distribution evolves very little with time. For clarity, we therefore follow the
procedure of Ref.~\cite{Burnier:2011bf} and we subtract from
the time-evolved vector charge distribution $n_V$ the vector charge distribution without 
chiral magnetic wave, that is the distribution $n_V\vert_{B=0}$ obtained from evolving
Fig.~\ref{fig2} in the absence of a magnetic field. In Fig.~\ref{fig4}, we show this difference
normalized to the time-dependent vector charge density at the center $x=0$. The top
panel of Fig.~\ref{fig4} shows how the chiral magnetic wave separates vector charge density
in the transverse plane with increasing time. The middle and lower panel 
of Fig.~\ref{fig4} shows that our analytical solution (\ref{eq31})-(\ref{eq33}), although being 
accurate only to leading order in $\mu/T$, accounts for the full numerical result of the normalized
charge difference $\left(n_V- n_V\vert_{B=0}\right) /n_V\vert_{B=0}$ very well. To this end,
we show first that analytical and numerical results for this ratio coincide almost for vanishing
shear viscosity, and that a finite value of $\eta/s$ reduces the effect only mildly. 
 We then show in the lower panel that the analytical result is highly sensitive to the value
of the transverse velocities $v_\pm$ that determine the norm of the spatial separation $s_\pm$ 
in (\ref{eq32}). This demonstrates that equation (\ref{eq32}) and the spatial charge separation
plotted in Fig.~\ref{fig2} provides a quantitative basis for understanding the size of the effect. 
Fig.~\ref{fig5} then shows how the corresponding asymmetry of the axial charge distribution
$n_A$ develops with time. Since this distribution vanishes at $\tau_0$ for the initial condition
of Fig.~\ref{fig3} studied here, we normalize the distribution to $n_V\vert_{B=0}$.
%

Distributions for $ n_V- n_V\vert_{B=0}$ and $n_A$ at times $\tau=10$ fm were shown
previously in Figs.1 and 2 of Ref.~\cite{Burnier:2011bf}. In contrast to the present study, these
results were obtained for a static medium and a constant time-independent magnetic field 
of strength $e\, B = m_\pi^2$ which corresponds approximately to the peak value of 
$eB_x(\tau)$ in Fig.~\ref{fig1}. Also, this study in Ref.~\cite{Burnier:2011bf}  was for a 
smaller constant temperature $T = 165$ MeV which according to our discussion helps to
maximize the effect. Ref.~\cite{Burnier:2011bf} set the initial axial chemical potential to
zero, consistent with the initial conditions considered here. They initialized a two-dimensional 
vector chemical potential with a Wood-Saxon profile and they evolved  the chiral magnetic 
wave equation (\ref{eq22}) numerically. Despite these qualitative and quantitative differences, 
we observe that if one cuts the resulting two-dimensional distributions 
$ n_V- n_V\vert_{B=0}$ and $n_A$  of Ref.~\cite{Burnier:2011bf} at transverse position $y=0$
along the direction of the magnetic field, their shape and the transverse positions of their peaks
resembles closely the results shown in Fig.~\ref{fig4} and Fig.~\ref{fig5} of the present work.  
%

Given that the transverse displacement $s_+ - s_-$ of charges is very small (see Fig.~\ref{fig2}),
it is clear that the positions of the peaks of the distributions in 
Fig.~\ref{fig4} and Fig.~\ref{fig5} cannot be of dynamical origin alone. 
Rather, irrespective of the exact size of $v_\pm$ and of the
time-dependence of $s_+ - s_-$ that is sensitive to the chiral magnetic wave, the peaks in the 
distribution of Fig.~\ref{fig4} and Fig.~\ref{fig5}  will always be positioned at approximately twice 
the Gaussian width of $n_V(\tau_0)$, since they result from subtracting two equally normalized
Gaussian distributions with almost same width. This conclusion may be expected to depend
only mildly on whether the shape of the charge distribution is Gaussian or Wood-Saxon, and 
this may explain the similarity between our results and those of Ref.~\cite{Burnier:2011bf}.
As for the size of the effect, we note that the results of Fig.~\ref{fig4} and ~\ref{fig5} are normalized 
such that they are independent of the absolute value of the charge density. According to our results,
the effective charge transport caused by the chiral magnetic wave affects only a very small fraction
$O(10^{-9} - 10^{-10})$ of all the vector charges at mid-rapidity. Similarly, Fig.~\ref{fig5} indicates
that the dynamical build-up of an axial charge asymmetry is a numerically very small effect, at 
least for the input parameters considered here. One may wonder whether one can obtain 
numerically larger effects by changing input parameters. Given that the size of the effect is
clearly governed by the integral (\ref{eq32}), the only way of maximizing the effect is to go to
even larger magnetic fields and/or to even smaller temperatures than those estimated for 
collisions at SPS energy. The results found for upper RHIC (LHC) energy are similar in shape to 
those in Figs.~\ref{fig4}, ~\ref{fig5}, but they are approximately one (two) orders of magnitude
smaller in size (data not shown). 
Here, we note that even if we choose a time-independent
magnetic $e\, B = m_\pi^2$, the distribution in Fig.~\ref{fig4} does not rise larger than
peak values of $O(10^{-5})$ which is still less than one charged particle per collision.

\section{Initial conditions with finite axial chemical potential}
\label{sec4}
For a fluid dynamic evolution initialized with vanishing axial charge density, we have seen
in section~\ref{sec3} that the chiral magnetic wave is unlikely to generate an experimentally 
testable electric quadrupole moment within the mesoscopic and finite-lived systems created in 
ultra-relativistic heavy ion collisions. In our studies, the fraction of the total charge density, 
separated by the CMW along the direction of $B_x$ turned out to be several orders of 
magnitudes too small, even under optimistic choices of thermodynamic variables, see 
Fig.~\ref{fig4}. We have traced back this finding to the parametrically and numerically small 
value of the transverse velocity $v_x$ of charge transport, which leads to spatial separations 
of vector charge that are typically much smaller than 1 fm, see Fig.~\ref{fig2}. 
%
\begin{figure}[h]
\begin{center}
\includegraphics[width=8.cm]{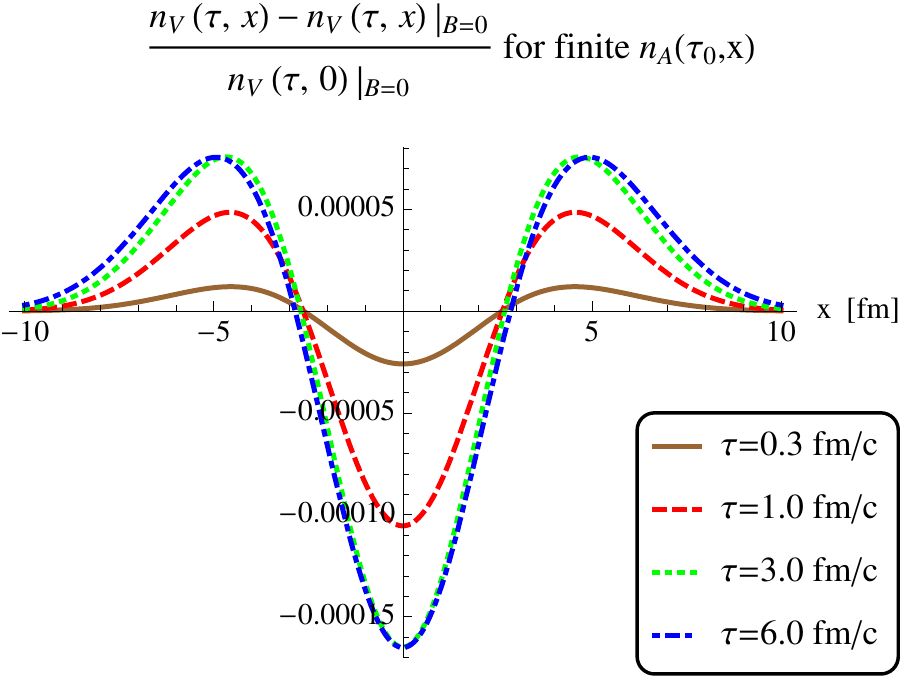}\\
\vspace{.5cm}
\includegraphics[width=8.cm]{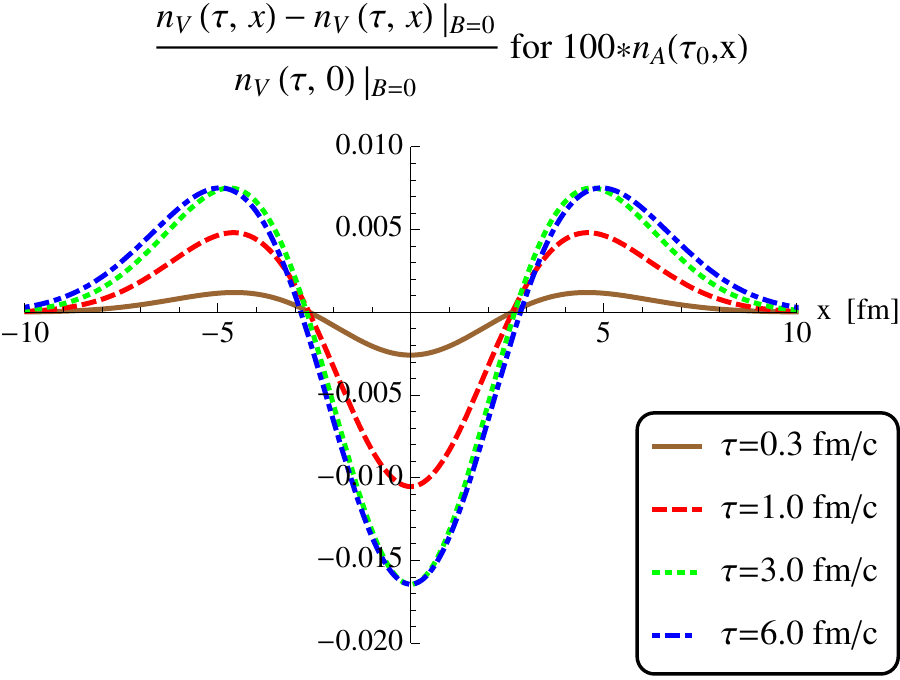}\\
\vspace{.5cm}
\includegraphics[width=8.cm]{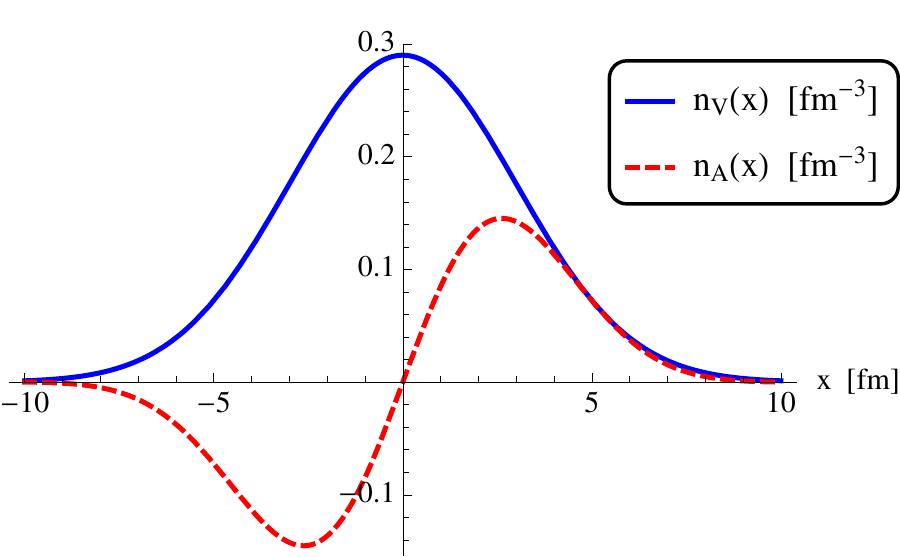}
\end{center}
\caption{Top and middle panel: same as in the top panel of Fig.~\ref{fig4}, but calculated
for the shown non-vanishing initial axial charge density distribution $n_A$ (top panel), and
for a 100 times larger density $n_A$ (middle panel). 
}\label{fig6}
\end{figure}

 We now investigate to what extent this conclusion can change for initial conditions with 
 non-vanishing axial charge asymmetry in the direction of $B_x$. The physical
 motivation for exploring this possibility is two-fold. First, for a non-vanishing initial axial charge 
 density distribution as the one shown in the lower panel of Fig.~\ref{fig6}, vector charge separation 
can be expected to be dynamically more efficient since a sizable finite axial chemical potential  
needed for vector charge transport is present from $\tau_0$ onwards and does not need to be
generated dynamically from gradients of the vector charge distribution. Second, as pointed out
in particular in the recent work of Ref.~\cite{Hongo:2013cqa}, the non-conservation
of the axial current $\partial_{\mu} j_A^{\mu} = C\, E.B$ may provide a source for a spatial
asymmetry of $n_A(x)$ along the direction of $B_x$. This is so, if one starts from a picture of
the heavy ion collision in which there is not only a strong $B_x$-field, but also a strong electric 
field $E$, and if the $E$-field is directed preferentially along (opposite) $B_x$ in the positive 
(negative) transverse half-plane $x> 0$ ($x<0$), as one expects from general geometric
considerations. In the study of Ref.~\cite{Hongo:2013cqa}, 
such a term was included in the dynamical evolution. 

We recall that the chiral magnetic wave leads to spatial asymmetries of the axial charge in response to
a finite vector charge chemical potential, and it does not require a finite $E$-field. In contrast,
a creation of spatial asymmetries in $n_A$ from   $\partial_{\mu} j_A^{\mu} = C\, E.B$ requires a
finite $E$-field but does not depend on a finite vector charge chemical potential. Here, 
we do not address the important question about the size of the spatial asymmetries in $n_A$
that may be created by this alternative mechanism, nor do we discuss the time scale on which 
such an asymmetry may arise in the collision. Rather, we simply accept the logical possibility
that a finite spatial asymmetry of $n_A$ along $B_x$ may be present in the initial conditions and
we ask how efficient the CMW could be in transfering this spatial asymmetries in $n_A$ into 
a spatial separation of the vector charge distribution.
To this end, we have repeated in Fig.~\ref{fig6} the calculation shown
in the upper panel of Fig.~\ref{fig4} for the same input values and the same initial vector charge
density $n_V(\tau_0,x)$, but for the non-vanishing axial charge density $n_A(\tau_0,x)$ shown in 
Fig.~\ref{fig6}. We find that - although the signal is still numerically small - it is enhanced by
almost 5 orders of magnitude compared to the case in Fig.~\ref{fig4}. This supports the general
argument that the mechanism of vector charge separation due to the chiral 
magnetic wave is significantly more efficient if a finite axial chemical potential needs not be 
generated dynamically but is present in the initial conditions. We further observe that for a
given $x$-dependence of the initial asymmetry $n_A(\tau_0,x)$, the size of the vector charge 
separation grows approximately linearly with the amplitude of $n_A(\tau_0,x)$. In fact, rescaling
$n_A$ in Fig.~\ref{fig6} by a factor 100 (which for the parameters chosen here still corresponds
to $\mu_A(\tau,x)/T(\tau)  < 1$ everywhere), we find that the relative size of the normalized
vector charge difference $\left(n_V- n_V\vert_{B=0}\right) /n_V\vert_{B=0}$ increases by a 
factor $\sim 100$ as well. We note as a speculative aside that if one starts from the numerical
finding that the peak values of $eB_x$ reached in heavy ion collisions are $O(m_\pi^2)$ and
{\it if} one assumes that 
$\vert eE\vert \sim \vert eB_x \vert \sim O(m_\pi^2)$, then the non-conservation of the 
axial current is parametrically large, $\partial_\mu j_A^{\mu} \propto \textstyle{\frac{m_\pi^4}{e}}$ 
and large initial values for $n_A$ may be feasible indeed. 

\section{Conclusion}
The chiral magnetic wave is a direct and unavoidable consequence of formulating QCD
hydrodynamics in an external electromagnetic field. Experimental confirmation of this 
phenomenon would be of considerable interest since it would underline the relevance 
of quantum anomalies for the properties of hot and dense matter. However, whether
the chiral magnetic wave can lead to unambiguous signals of experimentally accessible 
strength in ultra-relativistic heavy ion collisions (and whether the CMW can offer a
dynamical explanation for experimental signals of given strength) is a question that 
requires theoretical exploration. In the present paper, we have shown that in an expanding
fluid and for small charge densities, the chiral magnetic wave separates axial charges with
a local velocity
\begin{equation}
  v_{\pm}(\tau) = \frac{C}{4C_1} \frac{B_x(\tau)}{T^2(\tau)}\, ,
  \label{eq38}
\end{equation}
where $C = \textstyle\frac{e}{2\pi^2}$ is the strength of the axial anomaly and $C_1$ is a 
factor of order $O(1)$ in the equation of state (\ref{eq25}). For the special case of a 
fluid that satisfies to leading order in $\mu/T$ the symmetries of Bjorken's model and
where propagating charges can be treated as perturbations at subleading order in  $\mu/T$,
we have derived explicitly the equations of motion for  the chiral magnetic wave and the
velocity (\ref{eq38}). Moreover, we have argued on general grounds that we expect 
equation (\ref{eq38}) to describe the CMW-induced velocity of charge separation also for 
fluids whose collective expansion differs from that of the Bjorken-type model studied in sections 
~\ref{sec3} and ~\ref{sec4}. 

For choices of the magnetic field strength and of the time-dependent temperature that are
realistic for ultra-relativistic heavy ion collisions, the values found for $v_x$ from (\ref{eq38})
are non-relativistic, and the resulting time-integrated value of the spatial separation of 
right- and left-handed charges tends to be very small, see Fig.~\ref{fig2}. For the simplified
Bjorken-type model studied in section~\ref{sec3}, the CMW therefore typically induces
asymmetries in the electric charge distribution that we regard as being too small to be 
experimentally accessible since they affect on average the distribution of much less than one
charge per collision. This conclusion is reached on the basis of a simplified model.
While numerical results may change somewhat for the case
of a fully realistic fluid dynamic simulation of relativistic heavy ion collisions, we emphasize
that our main conclusion is based on the logarithmic order of the asymmetry of charge separation
($O(10^{-8})$ in Fig.~\ref{fig4}) - a more realistic simulation alone is unlikely to overcome this
big suppression factor. As discussed in section~\ref{sec4}, however, there may be physics effects 
that are different from the chiral magnetic wave and that enhance its phenomenological 
consequences. The findings of section ~\ref{sec4} therefore call for a better understanding of the initial conditions with which the CMW is initialized.

 \noindent
 {\it Acknowledgements.}
One of us (S.F.T.) would like to thank the CERN TH Unit for hospitality during this work. 
We thank  Navid Abbasi, Mohammad Ali-Akbari, Farhad Ardalan, Stefan Fl\"orchinger, 
Massimo Giovannini, Larry McLerran and Sergey Voloshin for useful discussions. 
 

\end{document}